\documentclass[%
reprint,
superscriptaddress,
nobibnotes,
 amsmath,amssymb,
 notitlepage, 
 twocolumn,
 prb,
]{revtex4-2}

\usepackage{pdfpages}
\usepackage{etoolbox} 

\usepackage{graphicx}
\usepackage{dcolumn} 
\usepackage{bm} 
\usepackage{multirow}
\usepackage{mathptmx}
\usepackage{siunitx}
\usepackage{color}
\usepackage[caption=false]{subfig}

\graphicspath{{./img/}}

\newcommand{\beq}[1]{\begin{equation}\label{#1}}
\newcommand{\eep}{\;.\end{equation}}
\newcommand{\eec}{\;,\end{equation}}
\newcommand{\eeq}{\end{equation}}

\newcommand*\dd{\mathop{}\!\mathrm{d}} 

\newcommand{\grad}{\nabla}

\newcommand{\lb}{\left(}
\newcommand{\rb}{\right)}

\newcommand*\chem[1]{\ensuremath{\mathrm{#1}}} 

\renewcommand{\a}{\alpha}

\renewcommand{\d}{\delta}
\newcommand{\ep}{\epsilon}
\newcommand{\la}{\lambda}
\renewcommand{\th}{\theta}

\newcommand{\p}{\phi}

\newcommand{\D}{\Delta}

\newcommand{\Om}{\Omega}

\DeclareMathAlphabet{\mathcal}{OMS}{cmsy}{m}{n} 

\newcommand{\Ef}{\mathcal{E}}   

\newcommand{\V}{\mathcal{V}} 

\newcommand{\Vel}{\mathcal{V}_{\text{elastic}}}
\newcommand{\Velec}{\mathcal{V}_{\text{elec}}} 
\newcommand{\Vstack}{\mathcal{V}_{\text{stack}}} 
\newcommand{\Vtot}{\mathcal{V}_{\text{tot}}}

\newcommand{\Nmax}{N_{\text{max}}}

\newcommand{\Asc}{A_{\text{M}}} 
\newcommand{\As}{A_{\text{s}}} 

\newcommand{\Ct}{\mathcal{C}_3}
\newcommand{\Cs}{\mathcal{C}_6}

\newcommand{\bvec}[1]{\mathbf{#1}}

\makeatletter
\patchcmd{\@outputpage@head}{\@ifx{\LS@rot\@undefined}{}{\LS@rot}}{}{}{}
\makeatother

\begin{document}

\title{Theory of polar domains in moir\'e heterostructures}

\author{Daniel Bennett}
\email{db729@cantab.ac.uk}
\affiliation{Theory of Condensed Matter, Cavendish Laboratory, Department of Physics, J J Thomson Avenue, Cambridge CB3 0HE, United Kingdom}
\affiliation{Physique Théorique des Matériaux, QMAT, CESAM, University of Liège, B-4000 Sart-Tilman, Belgium}

\date{\today}

\begin{abstract}

The discovery of ferroelectric behavior in twisted bilayers without inversion symmetry has prompted the consideration of some moir\'e heterostructures as polar materials. However, misconceptions about the nature and origin of the observed ferroelectricity indicate that a better theoretical understanding of the polar properties of moir\'e heterostructures is needed. In this paper, it is proposed that all moir\'e heterostructures in which there is a local breaking of inversion symmetry exhibit an out-of-plane moir\'e polar domain (MPD) structure, and is verified using first-principles calculations for several different bilayer systems. In transition metal dichalcogenide bilayers, a deformation of the charge density on each layer occurs due to the change in stacking arrangements throughout the moir\'e superlattice, leading to a local out-of-plane dipole moment, with the magnitude and shape of the MPDs being dominated by the chalcogen atoms. While the MPDs in all bilayers considered were found to be sensitive to the moir\'e period, it is only in the aligned homo-bilayers that they can be tuned with an out-of-plane electric field. The misconceptions about ferroelectricity in moir\'e heterostructures are addressed, and it is proposed that the only scenario in which the MPDs can be considered ferroelectric domains is via a global van der Waals sliding in a homo-bilayer. Finally, a general theoretical discussion of the polar properties of moir\'e heterostructures without inversion symmetry is provided.
\end{abstract}

\maketitle

\section{Introduction}

Since the discovery of superconducting \cite{cao2018unconventional,yankowitz2019tuning} and insulating \cite{cao2018correlated} behavior in magic angle graphene, twistronics has quickly become one of the most active fields in condensed matter physics. One of the most recent advancements in the field is the discovery of ferroelectric behavior in twisted hetero-bilayers \cite{zheng2020unconventional,yasuda2021stacking}. In addition to the observations of ferroelectricity, polarization domains have been observed without an overall ferroelectric response \cite{woods2021charge}. A switching of polarization in a stacking domain of hexagonal boron nitride (hBN) via van der Waals sliding has also been observed \cite{stern2020interfacial}. Although claimed to be ferroelectricity, insufficient measurements were taken to show hysteric behavior of the sample. In Ref.~\cite{yasuda2021stacking}, a first-order switching of the polarization domains was observed in \textit{untwisted} hBN, but in twisted hBN the hysteresis loop was continuous and very narrow. Thus, how and when moir\'e superlattices exhibit ferroelectric behavior is still unclear.

It had been proposed a few years before the discoveries that two-dimensional layered systems could exhibit a local out-of-plane polarization domain structure \cite{li2017binary}, which could be inverted by a relative `van der Waals' sliding between the layers, changing the stacking arrangments. Applying this concept to a twisted bilayer, the stacking domains can be considered polar domains, where neighboring domains have opposite polarization. It has been suggested that the relative sizes of these domains can be tuned with the application of an electric field perpendicular to the bilayer, leading to a nonzero total polarization of the superlattice \cite{bennett2022electrically,enaldiev2021scalable}, which can be very nonlinear when the moir\'e period is large. Although experimental measurements clearly show a ferroelectric response in homo-bilayers, from a theoretical perspective an ideal homo-bilayer has zero total polarization at zero field \cite{bennett2022electrically}. Even in an imperfect sample, which would have a nonzero total polarization due to uneven stacking domains, the orientation at zero field cannot be inverted. Thus, an ideal homo-bilayer does not satisfy the criteria for ferroelectricity via this mechanism, and additional effects must be required to explain the experimentally observed behavior. 

Separate to the question of whether or not the overall system exhibits a ferroelectric response is the question of whether or not the stacking domains themselves are ferroelectric. Several theoretical studies have claimed, without any justification, that the stacking domains in homo-bilayers are ferroelectric \cite{ferreira2021weak,enaldiev2021piezoelectric,enaldiev2021scalable}. This is simply incorrect: the domains are determined by the structure, and even if they change in area in response to an applied field, the orientation of the polarization in each domain is pinned cannot be inverted \cite{bennett2022electrically}.

It is possible that the domains in homo-bilayers, could become ferroelectric via van der Waals sliding. An electric field is applied locally with a biased tip which scans the sample, and the average polarization is calculated. It has been suggested that when the local polarization is anti-aligned with the field, a relative sliding occurs in order to change the stacking configuration and invert the polarization. However, in order for the domains to be ferroelectric, a global sliding of the layers must occur so that the orientation at zero field is reversible. Because the field is applied locally by scanning a biased tip over a sample, it is not clear whether the sliding occurs locally beneath the tip, and reverses once the tip moves on, or globally, and persists when the field is removed. Thus, while possible, it is not immediately clear whether the stacking domains are actually ferroelectric via this mechanism. Furthermore, in Ref.~\onlinecite{stern2020interfacial}, a switching of the orientation of a domain due to an applied field was observed, but whether or not this switching persisted at zero field was not investigated. In Ref.~\onlinecite{yasuda2021stacking}, a first-order polarization switching was observed in untwisted samples, but not twisted samples. Thus, ferroelectricity in moir\'e heterostructures via van der Waals sliding has not yet been experimentally observed.

In order to address the question of if and when moir\'e heterostructures exhibit ferroelectricity, and when the stacking domains can be considered ferroelectric domains rather than just polar domains, a better theoretical understanding of polar phenomena in moir\'e heterostructures is needed. One possible reason for the misconceptions mentioned above is that the discovery of polar properties of moir\'e heterostructures is relatively recent. They are very different from conventional ferroelectric materials such as oxide perovskites (\chem{ABO_3}), where polarization arises from soft-mode lattice instabilities, e.g.~the off-centering of the B cations with respect to the \chem{O_6} octahedra \cite{megaw1952origin,migoni1976origin}. For example, in barium titanate (\chem{BaTiO_3}, BTO) and lead titanate (\chem{PbTiO_3}, PTO), the Ti off-centering and ferroelectricity have been attributed to the hybridization of the titanium $3d$ and oxygen $2p$ states \cite{cohen1992origin}. The physical mechanism for polarization in moir\'e superlattices, the local breaking of out-of-plane inversion symmetry, is completely different, and while two unique phenomena have already been attributed to the domains, they are not well-understood in general. For example, while the polar domains in homo-bilayers (two identical monolayers, regardless of orientation) have been studied, the properties of polar domains in hetero-bilayers (two different monolayers), have not been investigated.

The aim of this paper is to propose that polar domains are a fundamental property of \textit{all} moir\'e heterostrctures which do not locally have an out-of-plane inversion symmetry, and to establish a basis for the theoretical understanding of these domains. In order to avoid further confusion with conventional ferroelectric domains, I will refer to polar domains in moir\'e heterostructures as `moir\'e polar domains' (MPDs) throughout this paper. The out-of-plane MPDs in bilayer hBN and the transition-metal dichalcogenides (TMDs) \chem{MX_2/M'X'_2}, with \chem{M,M'=Mo,W} and \chem{X,X'=S,Se}, are measured from first-principles calculations. It is demonstrated that the MPDs lead to non-uniform dielectric properties in moir\'e superlattices, such as interlayer charge transfer, polarizability, and the evolution of the local band gap due to an applied field. Although the concept of a local band gap is in general ill-defined, it has been used to estimate the potential landscape for interlayer moir\'e excitons \cite{jung2014ab,wu2014tunable,zhang2017interlayer,yu2017moire,wu2018theory,seyler2019signatures}, giving rise to novel exciton dynamical, optical and many-body phenomena \cite{remez2021dark}. While the theory of lattice relaxation and its influence on structural properties of moir\'e superlattices is well-known \cite{nam2017lattice,carr2018relaxation}, the practical details required to perform efficient calculations are typically omitted.  A complete study of lattice relaxation in moir\'e heterostructures is provided in the Appendix, with particular emphasis on the use of crystal symmetries to greatly reduce the computational effort of the calculations. The properties of MPDs in homo- and hetero-bilayers as a function of moir\'e period (twist angle) and electric field via lattice relaxation are then discussed. Finally, a discussion of MPDs as a fundamental property of moir\'e heterostructures without inversion symmetry is provided. While electrically tunable domains via lattice relaxation and first-order polarization switching via van der Waals sliding are interesting and unique properties, they are not criteria for moir\'e heterostructures to be considered polar materials. In light of the understanding of MPDs established in this paper, the question of whether moir\'e heterostructures are ferroelectric, and how they differ from conventional ferroelectrics, is addressed.

\section{Methods}
\subsection{Theoretical model}

The model originally used to describe the polar response in homo-bilayers via electrically tunable lattice relaxation \cite{bennett2022electrically} is used to describe MPDs in a general moir\'e heterostructure. It is a continuum elastic model, described by a total free energy which is an integral of a local energy density over a moir\'e period:
\beq{eq:V_tot_main}
V_{\text{tot}} = \frac{1}{\Asc}\int_{\Asc} \V_{\text{tot}}(\bvec{r}+\bvec{U}(\bvec{r}))\dd\bvec{r}
\eec
where $\Asc$ is the area of the moir\'e period, and a displacement field $\bvec{U}(\bvec{r})$ is included in order to allow for lattice relaxation. This model is general, and can describe various phenomena in moir\'e superlattices at the continuum level, depending on the contributions included in $\Vtot$. Typically, the intralayer elastic energy and interlayer stacking energy are included, and Eq.~\eqref{eq:V_tot_main} is minimized with respect to $\bvec{U}(\bvec{r})$ in order to describe the reconstruction of the stacking domains \cite{jung2015origin,nam2017lattice,zhang2018structural,carr2018relaxation}. For large moir\'e periods (small twist angles), the effect of lattice relaxation becomes significant, and leads to wide stacking domains separated by sharp domain walls. By including the electrostatic energy due to the coupling between the MPD and an applied field, the effect of an applied field on lattice relaxation can be described \cite{bennett2022electrically}. The total energy density is then given by the sum of elastic, stacking and electrostatic energy densities:
\beq{eq:V_tot_all}
\begin{gathered}
\Vtot(\bvec{r}) = \V_{\text{elastic}}(\grad\bvec{U}(\bvec{r})) + \V_{\text{stack}}(\bvec{r}) + \Velec(\bvec{r})\\[5pt]
\Vel(\grad\bvec{U}(\bvec{r})) = C_{ijkl}\ep_{ij}\ep_{kl}\\
\Vstack(\bvec{r}) = \V_0(\bvec{r})\\
\V_{\text{elec}}(\bvec{r}) = - \Ef p_0(\bvec{r})
\end{gathered}
\eeq
where $\ep_{ij} =\frac{1}{2}\lb \partial_i U_j + \partial_j U_i\rb$ is the strain tensor, $C$ is the elastic tensor \cite{carr2018relaxation,zhu2020twisted,bennett2022electrically}, $\V_0$ is the cohesive energy, $\Ef$ is component of the applied electric field perpendicular to the bilayer and $p_0(\bvec{r})$ is the local out-of-plane dipole moment. It is assumed that the layer separation $d(\bvec{r})$ is at its equilibrium value $d_0(\bvec{r})$ everywhere in the supercell, neglecting the effects of higher order elastic contributions \cite{jung2015origin} and the dielectric response to the field \cite{bennett2022electrically}. For a given twist angle, the local energy densities in Eq.~\eqref{eq:V_tot_all} can be parameterized using first-principles calculations at zero strain. The strain, and hence the structure, can then approximated by minimizing Eq.~\eqref{eq:V_tot_main}. A better way to approximate the structure would be to perform a geometry relaxation from first-principles, but this would be extremely expensive for large supercells.

Even without performing geometry relaxations, parameterizing Eq.~\eqref{eq:V_tot_all} can be difficult and expensive. Expensive calculations can be avoided by using the mapping from real space to configuration space \cite{carr2018relaxation,bennett2022electrically}, where the local configurations in real space are condensed into a single primitive cell:
\beq{eq:slide_map}
\bvec{s}(\bvec{r}) = \lb I - R_{\th}^{-1} \rb \bvec{r} \mod \{ \bvec{a}_1,\bvec{a}_{2}\}
\eec
where $\bvec{s}$ is the corresponding position in configuration space, $R_{\th}$ is a rotation matrix, and $\bvec{a}_i$ are the lattice vectors of a commensurate bilayer. Although this mapping is exact, for angles which are small deviations from the commensurate stackings 3R ($\th = 0 + \frac{2n\pi}{3}$) and 2H ($\th=\frac{\pi}{3}+ \frac{2n\pi}{3}$), the rotation matrix can be Taylor expanded, and we find that the configurations are well approximated by the commensurate 3R or 2H stackings plus a relative translation between the layers. After mapping Eqs.~\eqref{eq:V_tot_main} and \eqref{eq:V_tot_all} to configuration space, Eq.~\eqref{eq:V_tot_all} can be parameterized with first-principles calculations using a commensurate bilayer and sliding one layer over the other, but only for twist angles close to the commensurate 3R and 2H stackings.

\subsection{First-principles calculations}

The structural properties and MPDs of bilayer hBN, four TMD homo-bilayers (\chem{MoS_2}, \chem{MoSe_2}, \chem{WS_2}, \chem{WSe_2}) and the six hetero-bilayers formed from different combintations of the aforementioned TMD monolayers were measured in configuration space using first-principles calculations, following the methodology in Ref.~\onlinecite{bennett2022electrically}. Density functional theory (DFT) calculations were performed using the {\sc siesta} method and code \cite{siesta}, using PSML \cite{psml} norm-conserving \cite{norm_conserving} pseudopotentials, obtained from pseudo-dojo \cite{pseudodojo}. {\sc siesta} employs a basis set of numerical atomic orbitals (NAOs) \cite{siesta,siesta_2}, and double-$\zeta$ polarized (DZP) orbitals were used for all calculations. The basis sets were optimized by hand, following the methodology in Ref.~\onlinecite{basis_water}.

A mesh cutoff of $1200 \ \text{Ry}$ was used for the real space grid in all calculations. A Monkhorst-Pack $k$-point grid \cite{mp} of $12 \times 12 \times 1$ was used for the initial geometry relaxations, and a mesh of $18 \times 18 \times 1$ was used to calculate the out-of-plane dipole moments. Calculations were converged until the relative changes in the Hamiltonian and density matrix were both less than $10^{-6}$. For the geometry relaxations, the atomic positions were fixed in the in-plane directions, and the vertical positions and in-plane stresses were allowed to relax until the force on each atom was less than $0.1 \ \text{meV/\AA}$. The layer separation $d$ was taken to be the distance between the two layers in bilayer hBN and the distance between the metals in the TMD bilayers, and the volume of the bilayers was taken to be $\Om = Ad$, where $A$ is the in-plane area of the unit cell. The cohesive energy per unit cell is taken to be $\V_0 = \V_{\text{bilayer}} - 2\V_{\text{mono}}$, where $\V_{\text{bilayer}}$ and $\V_{\text{mono}}$ are the total energies of the bilayer and monolayers, respectively. The polarizabilities of the bilayers, defined in Ref.~\cite{bennett2022electrically}, were not included in Eq.~\eqref{eq:V_tot_all}, but were calculated for completeness. The polarizability to zeroth order in $d$, $\a_0$, was obtained by fixing the relaxed geometry and applying strong positive and negative fields, which is required for the cancellation of errors in the polarization at zero field. The polarizability to first order in $d$, $\a_1$, was obtained by changing the layer separation by $\pm 1 \%$ with respect to $d_0$ and measuring the relative change in the polarizability. When an out-of-plane electric field was applied, a dipole correction \cite{dipole_correction_1,dipole_correction_2,dipole_correction_3,dipole_correction_4} was used in the vacuum region to prevent dipole-dipole interactions between periodic images. \\

Calculations were repeated to parameterize $\V_0$, $d_0$, $\a_0$ and $\a_1$, as well as the dipole moment $p_0$ and out-of-plane polarization $P_0=\frac{p_0}{\Om}$, in configuration space. The data were then fitted to a Fourier expansion, taking advantage of the $\Ct$ rotation symmetry of the bilayers. For a general scalar field $\p$, the following Fourier expansion was used:
\beq{eq:fourier_fitting}\resizebox{0.91\columnwidth}{!}{$
\begin{gathered}
\p(x,y) = \p_0 + \sum_{i=1}^{3} \p_i^{\text{even}}f_i^{\text{even}}(x,y) + \p_i^{\text{odd}}f_i^{\text{odd}}(x,y) \\[5pt]
f_1^{\text{even}}(x,y) = \cos{(2\pi x)} + \cos{(2\pi y)} + \cos{(2\pi (x+y))} \\
f_2^{\text{even}}(x,y) = \cos{(2\pi (x-y))} + \cos{(2\pi (2x + y))} + \cos{(2\pi (x+2y))} \\
f_3^{\text{even}}(x,y) = \cos{(4\pi x)} + \cos{(4\pi y))} + \cos{(4\pi (x+y))} \\[5pt]
f_1^{\text{odd}}(x,y) = \sin{(2\pi x)} + \sin{(2\pi y)} - \sin{(2\pi (x+y))} \\
f_2^{\text{odd}}(x,y) = \sin{(2\pi (y-x)} + \sin{(2\pi (2x+y))} - \sin{(2\pi (x+2y))} \\
f_3^{\text{odd}}(x,y) =  \sin{(4\pi x)} + \sin{(4\pi y)} - \sin{(4\pi (x+y))}\\
\end{gathered}
$}
\eeq
where $x$ and $y$ describe the configuration in units of the configuration space lattice vectors: $\bvec{s} = x \bvec{a}_1 + y \bvec{a}_2$. The first-principles calculations were performed along the line $(x,y) = (s,s)$, which includes all of the high symmetry stacking configurations, greatly reducing the number of calculations needed. However, it should be noted that $f_2^{\text{odd}}(s,s)=0$ for every value of $s$, so in order to be more precise, a full 2D parameterization of configuration space should be performed when considering properties which have higher-order odd components. Plots of the aforementioned quantities along the configuration space diagonal as well as the corresponding parameterization using Eq.~\eqref{eq:fourier_fitting} for the 11 bilayers considered can be found in the supplementary information (SI) \cite{SI}. Tables of the fitting parameters are also given.

Several electronic properties were also measured as a function of electric field in configuration space for one homo-bilayer (\chem{MoS_2}), and one hetero-bilayer (\chem{MoS2/MoSe_2}). The `local band gap' was determined from the electronic band structure at each value of $s$. The average charge density in the out-of-plane direction was obtained with {\sc c2x} \cite{c2x} using macroscopic averaging techniques \cite{junquera2003first,nanosmooth,stengel2011band,c2x}:
\beq{}
\rho_{\text{av}}(z) = \frac{1}{A}\int_A \rho(x,y,z) \dd{A}
\eec
on a grid of $10^{6}$ points. The change in the charge on the top layer is given by
\beq{}
\D q = A\int_{d_0/2}^{(d_0+c)/2} \rho_{\text{av}}(z) \dd{z}  - q
\eec
where $c$ is the out-of-plane lattice vector, and $q$ is the formal charge of the layer.

\subsection{Lattice Relaxation}

Following the methodology described in the Appendix, lattice relaxation calculations were performed using the parameterizations obtained from first-principles and $B = 49.866$ eV/\AA$^2$, $\mu = 31.548$ eV/\AA$^2$ for the bulk modulus and shear modulus respectively, from Ref.~\onlinecite{carr2018relaxation}. 3R-stacked \chem{MoS_2} and both stackings of \chem{MoS_2/MoSe_2} were considered, for twist angles ranging from $\th=1.0^{\circ}$ to $\th=0.1^{\circ}$ in steps of $0.1^{\circ}$, and electric fields ranging from $\Ef=-2$ V/\AA \ to $\Ef = +2$ V/\AA \ in steps of 0.5 V/\AA. The displacement fields were expanded in a Fourier series, using the following truncations in the Brillouin zone (BZ): five shells (10 $\bvec{k}$-vectors) for $\th \geq 0.5^{\circ}$, six shells (21 $\bvec{k}$-vectors) for $0.5^{\circ} > \th \geq 0.3^{\circ}$ and and seven shells (28 $\bvec{k}$-vectors) for $0.3^{\circ} >\th \geq 0.1^{\circ}$. The total energy was optimized with respect to the displacement field using the {\sc Optim} package in {\sc Julia}, using the analytic expression for the gradient of the free energy derived in the Appendix. The total energy was optimized using the limited memory BFGS algorithm (L-BFGS), with a tolerance of $1\times 10^{-5}$ eV. The elastic part of the energy and gradient are given exactly by Eqs.~\eqref{eq:V_elastic_exact} and \eqref{eq:G_elastic}, respectively, and the rest of the energy and gradient were obtained by numerical integration using the {\sc Cubature} package. The integrations could be parallelized, and the individual components of of the gradient could also be calculated in parallel, which could be employed in calculations involving a very large number of shells. In this study, since a large number of calculations were performed for different materials and different values of $\th$ and $\Ef$, separate calculations were executed in parallel, each using a single core. Calculations with the same value of $\th$ but different values of $\Ef$ were chained together by using the output of calculations at $\Ef=0$ as the starting point for calculations at $\Ef= \pm 0.5$ \ V/\AA \, and so on, reducing the number of iterations in each calculation considerably. The same procedure was also used to chain together calculations at zero field but at different twist angles, leaving additional shells (if any) empty for the first iteration.

\section{Results}

\begin{figure}[h!]
\centering
\includegraphics[width=0.9\columnwidth]{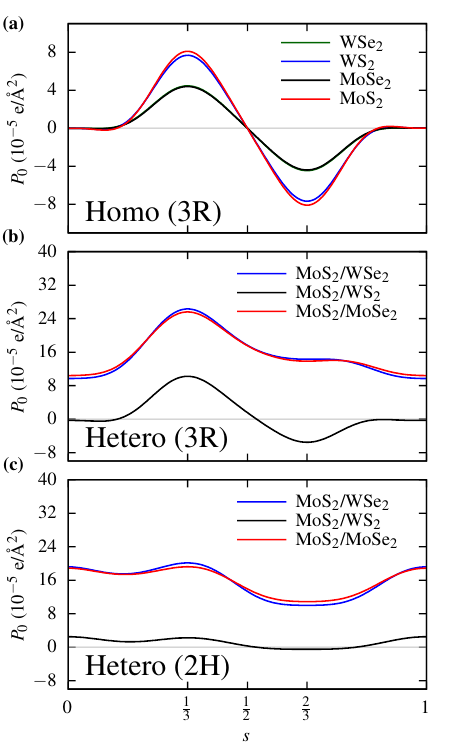}
\caption{Polarization along the configuration space diagonal for \textbf{(a)}: homo-bilayers (3R stacking), \textbf{(b)}: hetero-bilayers (3R stacking), and \textbf{(c)}: hetero-bilayers (2H stacking). For the hetero-bilayers, one example of each of the following is shown: same metal, different chalcogen (\chem{MoS_2/MoSe_2}), same chalcogen, different metal (\chem{MoS_2/WS_2}), different metal, different chalcogen (\chem{MoS_2/WSe_2}), with the rest available in the SI \cite{SI}.}
\label{fig:pol}
\end{figure}

\begin{figure*}[p!]
\centering
\includegraphics[width=\linewidth]{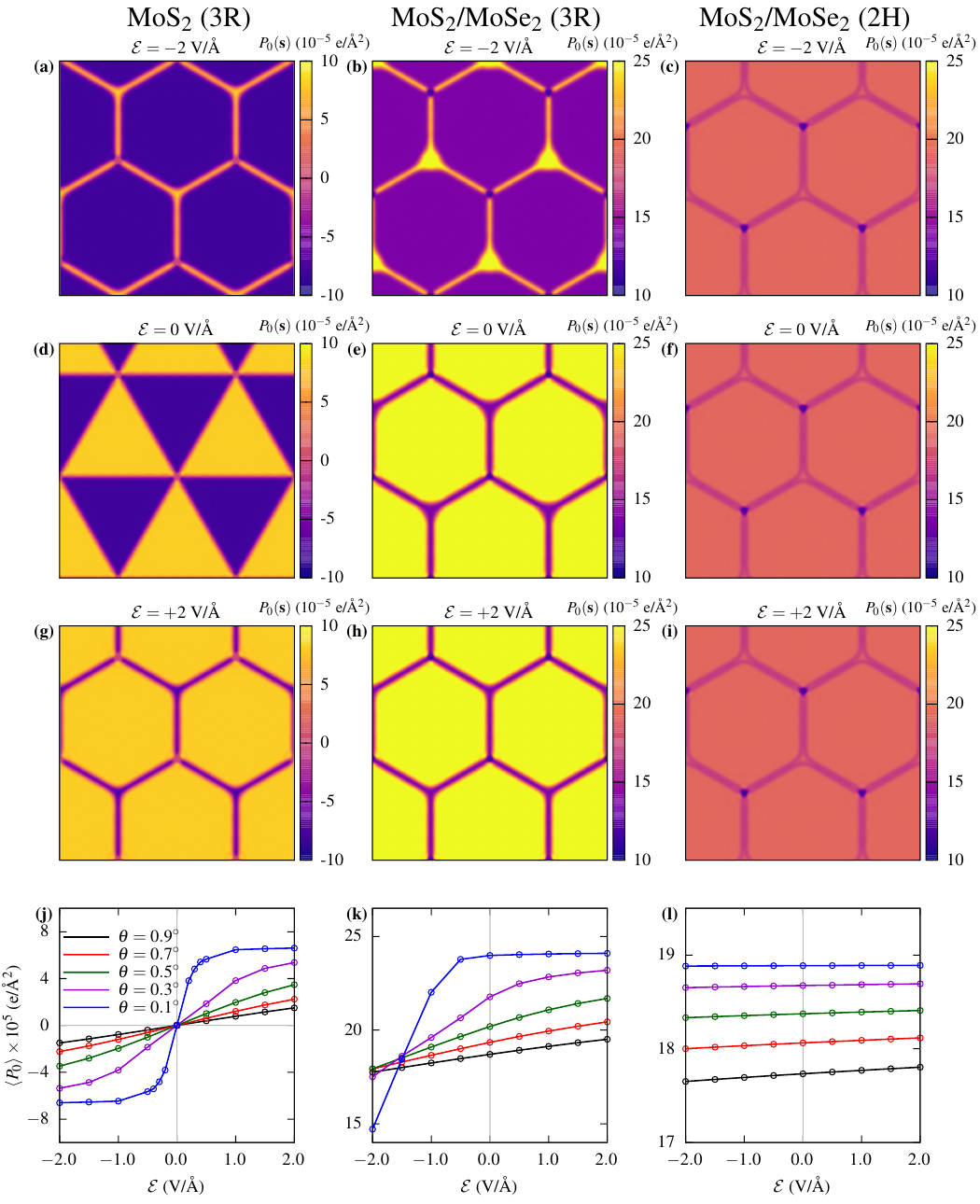}
\caption{MPDs for \textbf{(a)},\textbf{(d)},\textbf{(g)}: 3R-stacked \chem{MoS_2}, \textbf{(b)},\textbf{(e)},\textbf{(h)}: 3R-stacked \chem{MoS_2/MoSe_2} and\textbf{(c)},\textbf{(f)},\textbf{(i)}: 2H-stacked \chem{MoS_2/MoSe_2}, all at a twist angle of $\th = 0.1^{\circ}$ and electric field strengths of $\Ef = -2,0,2$ V/\AA. \textbf{(j)}-\textbf{(l)}: average polarization in configuration space as a function of $\Ef$ for several twist angles for each of the three bilayers.}
\label{fig:pol-contour}
\end{figure*}

The out-of-plane polarization of the TMD bilayers along the configuration space diagonal is shown in Fig.~\ref{fig:pol}. All of the homo-bilayers have an odd polarizaiton along the diagonal, with the magnitude being stronger in the sulfur TMDs. For hetero-bilayers with the same chalcogens but different metals, the MPD is nearly odd, but the polarization at the AB ($s=\frac{1}{3}$) and BA ($s=\frac{2}{3}$) stacking configurations are no longer equal in magnitude. When the chalcogens in each layer are different, the MPD has a very different form. The polarization is no longer centered about zero because the out-of-plane inversion symmetry is broken everywhere by having two different monolayers. For 2H-stacked homo-bilayers, as well as bilayers composed of elemental layers (e.g.~bilayer graphene), there is an out-of-plane inversion symmetry for every stacking configuration, and the polarization is zero everywhere. This is not the case for the  2H-stacked hetero-bilayers however, which also exhibit MPDs.

Lattice relaxation calculations were performed for one homo-bilayer (\chem{MoS_2}) and one hetero-bilayer (\chem{MoS_2/MoSe_2}) for a range of twist angles and electric fields. The results are summarized in Fig.~\ref{fig:pol-contour}, with additional plots available in the SI \cite{SI}. The behavior of the homo-bilayers is already known \cite{bennett2022electrically}: for a 2H stacking, there is no MPD and the structure is not influenced by an applied field. For a 3R stacking, any finite field breaks the $\Cs$ rotation symmetry of the model, causing the domains which are aligned with the field to grow, and the domains which are anti-aligned to shrink. For small angles and strong fields, the domain structure changes from a sharp triangular structure to a sharp hexagonal structure. For the hetero-bilayers, the shape of the MPDs was found not to be strongly influenced by an applied field. Changes in the 2H-stacked bilayers were negligible, and the 3R-stacked bilayers changed noticeably only for small twist angles and strong electric fields. Although Fig.~\ref{fig:pol-contour} (b) shows that the relative sizes of the domains change in 3R-stacked \chem{MoS_2/MoSe_2} for $\th=0.1^{\circ}$ and $\Ef = -2$ V/\AA, both domains have the same orientation, and thus the sign of the total polarization does not change.

The total polarization of \chem{MoS_2} (3R) and \chem{MoS_2/MoSe_2} (both stackings) as a function of $\th$ and $\Ef$ is summarized in Figs.~\ref{fig:pol-contour} (j)-(l). The 3R-stacked homo-bilayers show a nonlinear response to the field at small twist angles. The 3R-stacked hetero-bilayers only show a noticeable response when a strong field is applied in the direction opposite to the orientation of the MPDs, and the 2H-stacked hetero-bilayers do not show a noticeable response at any of the values of $\th$ and $\Ef$ used in this study. 

\begin{figure}[ht!]
\centering
\includegraphics[width=\columnwidth]{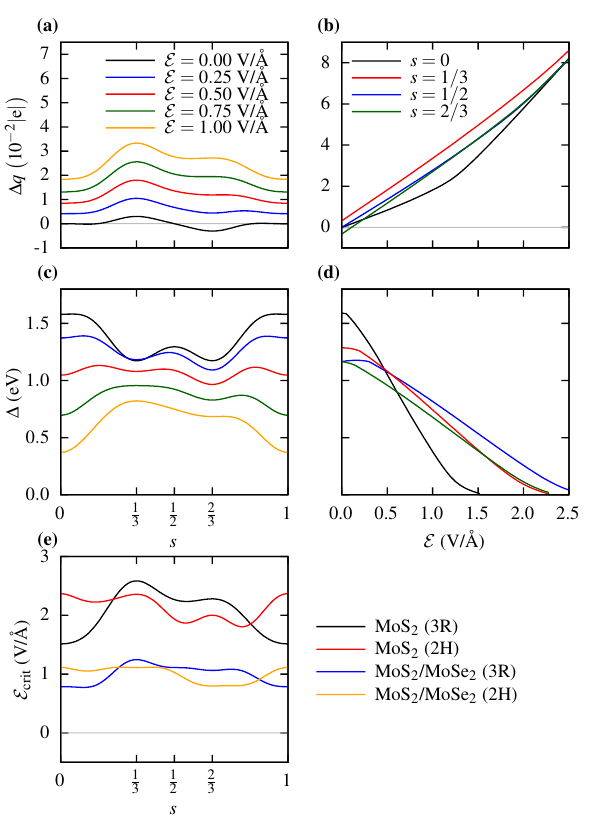}
\caption{Charge transfer $\D q$ for 3R-stacked \chem{MoS_2} as a function of \textbf{(a)}: $s$ along the diagonal through configuration space for several fixed values of $\Ef$ and \textbf{(b)}: $\Ef$, for several fixed values of $s$. \textbf{(c)} and \textbf{(d)}: the local band gap for the same values of $s$ and $\Ef$ as in \textbf{(a)} and \textbf{(b)}, respectively. \textbf{(e)}: Critical field $\Ef_{\text{crit}}$ required to close the local band gap in configuration space for both stackings of \chem{MoS_2} and \chem{MoS_2/MoSe_2}.}
\label{fig:gap-charge}
\end{figure}

The interlayer charge transfer, (the deformation of the electronic charge density on each layer, not the excitation of an electron across the layers), in 3R-stacked \chem{MoS_2} along the configuration space diagonal is shown in Fig.~\ref{fig:gap-charge} (a) for several values of $\Ef$. At zero field, we can see that the interlayer charge transfer exactly matches the shape of the MPD. For stronger fields, the charge transfer resembles the MPDs of the 3R-stacked hetero-bilayers with different chalcogens. The charge transfer as a function of field strength for the AA ($s=0$), AB, SP ($s=\frac{1}{2}$) and BA stacking configurations is shown in Fig.~\ref{fig:gap-charge} (b). It is continuous and linear for all except the AA stacking, which experiences a change in slope at $\Ef \approx 1$ \ V/\AA. These results are contrary to similar measurements in Ref.~\onlinecite{bennett2022electrically}, where the charge transfer was estimated from the M{\"u}lliken charges instead of the macroscopic charge density. No transfer was observed below $\Ef \approx 0.27$ V/\AA \, above which the charge transfer grew linearly with the field strength. The origin of the discontinuity was unknown, and because it was not observed here, may be an artifact of the previous calculations, since the M{\"u}lliken charges are not a well-defined quantity in a periodic solid.

Fig.~\ref{fig:gap-charge} (c) shows the local band gap in configuration space for different field strengths. It has been suggested that the excitonic properties of moir\'e superlattices may be tuned with an electric field, which would couple to the interlayer exciton dipole \cite{yu2017moire}. However, additional effects from the applied field, such as electrically tunable lattice relaxation, and the decreasing, and possibly closing of the local gap, have not been considered. Fig.~\ref{fig:gap-charge} (c) shows that the local band gap changes significantly, and non-uniformly due to the MPD, which should be accounted for when considering the possibility of tuning the excitonic properties of moir\'e superlattices. Fig.~\ref{fig:gap-charge} (d) shows the local gap as a function of electric field strength for the high symmetry stacking configurations. The AA configuration decreases most quickly and goes to zero at $\Ef \approx 1.5$\ V/\AA, which may explain the change in slope in the interlayer charge transfer. For the other configurations, the gap closes for $\Ef > 2$ \ V/\AA. Fig.~\ref{fig:gap-charge} (e) shows the critical field required to close the gap in configuration space for both stackings of \chem{MoS_2} and \chem{MoS_2/MoSe_2}. Although the local band gap measured in configuration space is in general not a physically meaningful quantity, Figs.~\ref{fig:gap-charge} (c) - (e) may provide some insight into the possibility of tuning the excitonic properties of moir\'e heterostructures at small twist angles with an applied field. Because electric fields are typically applied to real samples locally via a biased tip, knowledge of the nonuniform dielectric properties of moir\'e superlattices as a result of MPDs may provide some helpful insight in understanding experimental measurements.

\section{Discussion and Conclusions}

MPDs were found in all bilayers considered in this study, excluding 2H-stacked homo-bilayers, which have an out-of-plane inversion symmetry. The 3R-stacked Homo-bilayers considered exhibit MPDs which are odd and average to zero over the supercell. The MPDs in the hetero-bilayers are neither even nor odd, and do not average to zero over the supercell. For the TMDs, the magnitude and shape of the MPDs appears to be dominated by the chalcogens and is not strongly influenced by the transition metals. While the total polarization of the homo-bilayers (3R stacking) exhibits a nonlinear response to an applied field, the total polarization of the hetero-bilayers exhibits a negligible response in nearly all cases. The weak response to the field in the hetero-bilayers with different chalcogens is understandable: in the 3R-stacked homo-bilayers there is a $\Cs$ rotation symmetry of the stacking energy at zero field. Any finite electric field breaks this symmetry, leading to the uneven relaxation of the AB and BA domains. In the hetero-bilayers however, this symmetry is already broken by having two different monolayers, and thus there is already a difference in energy between the AB and BA domains. Applying a field does not lead to much additional splitting, and lattice relaxation is not significantly altered. All MPDs are sensitive to the moir\'e period, however.

While the behavior of the MPDs at angles which are small deviations from 2H and 3R stackings has been illustrated, the general behavior at all angles is far from clear. Although the configuration space mapping is valid for arbitrary twist angles, in principle we can only measure MPDs by performing a Taylor expansion of Eq.~\eqref{eq:slide_map} about a commensurate reference structure and parameterize systems at twist angles which are small deviations from the reference structure by sliding one layer over the other in first-principles calculations. It is not clear whether the parameterizations done using the commensurate reference structures would accurately describe the local polarization in bilayers which are at very large twist angles. This would need to be verified using computationally expensive first-principles calculations in real space.

However, it is an interesting theoretical question: consider a bilayer at an arbitrary twist angle $\theta$, which ranges from a reference structure at zero, for which the unit cell is commensurate, up to the first angle which leaves the system invariant. For hBN and the TMD bilayers, if we take the 3R stacking to be the reference structure at $\th=0$, then the 2H stacking is realized at $\th = \frac{\pi}{3}$, and the 3R stacking is realized again at $\th = \frac{2\pi}{3}$. For $\th = 0,\frac{\pi}{3},\pi,\ldots $ , the bilayer is commensurate and hence non-polar. In homo-bilayers, for small rational twist angles, we know that there is a MPD structure about the 3R stacking, and zero polarization about the 2H stacking. If we consider the twist angle as a tunable parameter, there should be a transition from a polar state to non-polar state between the 3R and 2H stackings. As a first approximation, if we were to parameterize the twist angles using the reference structure they are closest to, then we would expect a first-order polar-nonpolar transition exactly half-way between at $\th = \frac{\pi}{6}$. However, it is not clear whether the local polarization in bilayers at twist angles close to $\th = \frac{\pi}{6}$ (which would exhibit a quasi-crystal phase with 12-fold order \cite{moon2019quasicrystalline,yu2019dodecagonal,pezzini202030}) would be accurately described by parameterizations of configuration space using the 3R and 2H stackings. Therefore, a more robust way to locate and determine the order of the polar-nonpolar transition would be using first-principles calculations in real space.

\subsection{Are moir\'e heterostructures really ferroelectric?}

\subsubsection{Lattice relaxation}

To reiterate the conclusions of Ref.~\onlinecite{bennett2022electrically}, the total response of the spontaneous polarization arising from tunable lattice relaxation under an applied field does not qualify as ferroelectricity, since there is no remnant, switchable polarization at zero field. Despite this, ferroelectric behavior has been observed experimentally. The obvious difference between theory and experiment is that theoretical descriptions are of an ideal sample, whereas experimental samples are imperfect. Defects and strains induced by finite boundaries may lead to uneven domains, and a remnant polarization at zero field. Even so, this remnant polarization would not be switchable via lattice relaxation alone, because the system would relax back to its original configuration at zero field. The uneven motion of domain walls may lead to irreversible changes in the average polarization as an electric field is applied and removed. Perhaps electromechanical couplings such as piezoelectricity \cite{mcgilly2020visualization} and flexoelectricity \cite{mcgilly2020visualization,bennett2021flexoelectric,springolo2021direct} could also lead to irreversible structural changes as a field is applied and removed. It is easier to see that hetero-bilayers should not be ferroelectric, since the MPDs exhibit a negligible change in shape in response to an applied field. Even for fields strong enough to change the relative sizes of the domains, both types of domains have the same orientation but different magnitude, so it would be impossible to change the orientation of the total spontaneous polarization by applying a strong field in the opposite direction.

\subsubsection{van der Waals sliding}

A first-order polarization switching has been observed in hBN \cite{stern2020interfacial,yasuda2021stacking}. However, there are a number of conceptual problems which much be addressed before it can be determined if the observed behavior actually corresponds to a ferroelectric response. The images of the domains were made by scanning biased tips over the samples. It is thought that applying a strong electric field leads to a sliding by one third of a unit cell, switching the stacking configuration from e.g.~AB to BA, and aligning the local polarization with the field. Although claimed to be ferroelectric in this study and other theoretical studies \cite{wu2021sliding,zhong2021sliding}, whether or not this mechanism truly corresponds to ferroelectricity was not discussed. The field is applied locally, and it is not clear whether the sliding occurs locally, beneath the tip, and the atoms relax back to their initial configuration due to a large elastic energy penalty once the tip moves on, or whether a global sliding of one layer over the other occurs. The former corresponds to a local first-order switching of polarization, but not a ferroelectric response. The latter corresponds to a first-order switching of the entire domain structure, which in homo-bilayers would result in an inversion of the domains: $\bvec{P}(\bvec{r})\to -\bvec{P}(\bvec{r})$. For an ideal sample, this would still not correspond to ferroelectricity, since the polarization at zero field would always be zero, but for an imperfect sample, it is possible that the total polarization at zero field could be switched after sliding, and thus may qualify for ferroelectricity. In Ref.~\onlinecite{stern2020interfacial}, the orientation of the domains were measured at $\Ef = \pm 3$ V/\AA and a switching of the polarization was observed, although the orientation was not measured again after the fields were removed, so it is impossible to determine whether the orientation of the MPD at zero field was inverted or not. Thus, the claims of ferroelectricity were made without strong evidence. In Ref.~\onlinecite{yasuda2021stacking}, a first-order polarization switching was observed in untwisted hBN, but not in twisted hBN. Since neither study provides strong evidence of global van der Waals sliding in a twisted system, it is impossible to conclude whether moir\'e heterostructures can be ferroelectric, or whether MPDs can be considered ferroelectric domains via this mechanism.

\begin{figure}[h!]
\centering
\includegraphics[width=\columnwidth]{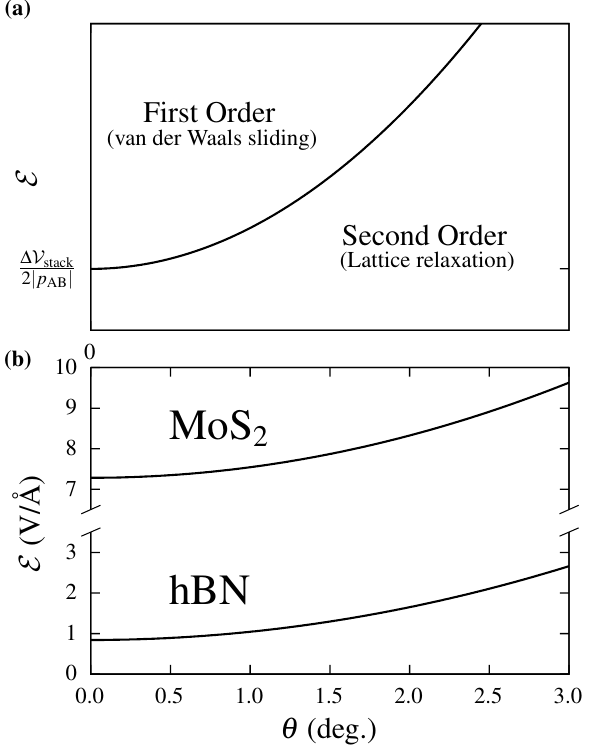}
\caption{\textbf{(a)}: Sketch of the phase boundary as a function of $\Ef$ and $\th$, separating a local first-order switching of the polarization via van der Waals sliding, and a second order change in the total polarization via lattice relaxation. \textbf{(b)}: parameterization of Eq.~\eqref{eq:first-order-2} for 3R-stacked hBN and \chem{MoS_2} using the values of $p_{\text{AB}}$ and $\D\Vstack$ calculated in this paper and $\mu_{\text{hBN}} = 43.6152$ eV/\AA$^2$ and $\mu_{\text{MoS}_2} = 31.548$ eV/\AA$^2$ from Refs.~\onlinecite{lin2019effective} and \onlinecite{carr2018relaxation}, respectively. }
\label{fig:first-order-phase}
\end{figure}

Using Eq.~\eqref{eq:V_tot_all}, we can obtain a rough estimate of when a first-order local sliding beneath a biased tip would occur in a twisted sample. Taking a 3R-stacked homo-bilayer for simplicity, we would expect a first order switching from AB $\to$ BA when $\Ef$ and $P_{\text{AB}}$ are anti-aligned, and when the energy gained from aligning the polarization with the field is greater than the cost of traversing the energy barrier, plus the elastic energy penalty associated with sliding:
\beq{eq:first-order-1}
\Ef \lb p_{\text{BA}} - p_{\text{AB}}\rb > \D\Vstack + \D\Vel
\eep
A sliding from AB to BA is achieved by a displacement $\bvec{u}(\bvec{x}_{\text{AB}}) = \begin{bmatrix} 1/3 \\ 1/3 \end{bmatrix}$, where $\bvec{x}$ is in units of $\bvec{a}_1$ and $\bvec{a}_2$ (see Appendix). From Eq.~\eqref{eq:V_elastic}, the elastic cost of such a displacement is $\D\Vel = \frac{4}{27}\mu\th^2$, and the energy barrier is taken to be the one at zero field for simplicity: $\D\Vstack = \V_{\text{SP}} - \V_{\text{AB}}$. Under these approximations, Eq.~\eqref{eq:first-order-1} predicts a phase boundary as a function of $\th$ for when a first-order sliding would occur in a given material:
\beq{eq:first-order-2}
\Ef_{\text{slide}} = \frac{1}{2\left| p_{\text{AB}}\right|} \lb \frac{4}{27}\mu\th^2 + \D\Vstack\rb
\eep
This is sketched in Fig.~\ref{fig:first-order-phase} (a). It should be noted that Eq.~\eqref{eq:first-order-1} is a very rough estimate, because it assumes that that the energy barrier is not reduced by the field, and only the point directly beneath the biased tip is displaced. In reality, a neighborhood around the tip would be displaced, although the precise behavior is difficult to estimate without a detailed study, which is beyond the scope of this paper.

We can also estimate the phase boundary for the materials considered in this paper. Fig.~\ref{fig:first-order-phase} (b) shows the parameterization of Eq.~\eqref{eq:first-order-2} for \chem{hBN} and \chem{MoS_2}. We can see that a first-order transition is certainly possible for hBN, although the field strengths required to achieve a sliding in \chem{MoS_2} are unrealistic. Even if the elastic energy penalty is neglected, the cost of traversing the energy barrier would require a field of $\Ef_{\text{slide}} > 7$ V/\AA.

\subsection{Comparison with conventional ferroelectrics}

\begin{table*}[ht!]
\renewcommand*{\arraystretch}{2}
\setlength{\tabcolsep}{10pt}
\begin{center}
\begin{tabular}{| c | c | c |}
\hline\hline
\textbf{System} & \textbf{Conventional ferroelectrics} & \textbf{Moir\'e heterostructures} \\ \hline
Prototypical materials & \chem{ABO_3} oxide perovskites (\chem{BaTiO_3}, \chem{PbTiO_3}, \ldots) & hBN, TMDs (\chem{MoS_2}, \chem{WS_2}, \ldots) \\ \hline
Mechanism for polarization & Soft mode lattice instabilities &  Local inversion symmetry breaking \\ \hline
Origin of polarization & Change in hybridization of B and O states &  Interlayer charge transfer\\ \hline
Origin of domains & Screen depolarizing field & Local inversion symmetry breaking \\ \hline \hline
\end{tabular}
\end{center}
\caption{Comparison of the polar properties of conventional ferroelectrics (oxide perovskites) and moir\'e heterostructures, assuming ideal structures.}
\label{table:ferro}
\end{table*}

The formation of ferroelectric domains, in oxide perovskite thin films for example, is a well-known problem which has been understood for many years. A free-standing ferroelectric thin film can form a polydomain structure in order to screen the depolarizing field arising from the polar discontinuities at the surfaces, and the behavior of the domains as a function of film thickness and perpendicular applied field can be studied in the limit of infinitely thin domain walls \cite{springer,vacuum_1,bennett2020electrostatics}. When the width of the domains $w$ is less than the thickness $d$ of the film, the domains follow Kittel's law \cite{springer,vacuum_1}: $w = \sqrt{l_{\text{k}}d}$, where $l_{\text{k}}$ Kittel's length, the relevant length scale for the domains. In the limit of zero thickness, the width of the domains diverge, and a monodomain or paraelectric phase becomes more favourable  \cite{bennett2020electrostatics}. Of course, in this limit, the approximation of infinitely thin walls is not appropriate. Ferroelectric domains with domain walls of finite width can be considered by generalizing to a Ginzburg-Landau theory \cite{kretschmer1979surface,chandra2007landau,luk2009universal}. At zero temperature, the free energy of a ferroelectric thin film can be given by
\beq{eq:GL_ferro}
V_{\text{tot}}^{\text{GL}} = \int \lb \la^2 \lb \nabla\bvec{P}(\bvec{r})\rb^2 + \frac{1}{4}\bvec{P}(\bvec{r})^4 - \frac{1}{2}\bvec{P}(\bvec{r})^2 - \Ef\cdot \bvec{P}(\bvec{r}) \rb \dd{\bvec{r}}
\eec
where $\la$ is the correlation length and $\bvec{P}$ is in units of the spontaneous polarization $\bvec{P}_0$ of the monodomain phase ($\la\to\infty$). Minimizing Eq.~\eqref{eq:GL_ferro} and solving the resulting equations, we find that the general solution for $\bvec{P}(\bvec{r})$ can be written in terms of elliptic functions \cite{luk2009universal}. At zero field, Eq.~\eqref{eq:GL_ferro} is clearly always invariant if we switch the orientation of the domains: $\bvec{P}(\bvec{r}) \to -\bvec{P}(\bvec{r})$, even in the monodomain and polydomain limits. This is not true of MPDs, although direct comparisons are difficult because in Eq.~\eqref{eq:V_tot_main} the polarization is not an order parameter, but rather it is determined by the structure.

It could be argued that the sign of the polarization in large regions of the moir\'e superlattice does change after relaxation when the domain walls move considerably. However, this argument relies on a poor definition of a polar domain. A domain can be defined as the area enclosed by a boundary on which an order parameter is zero. Although better definitions could be devised which account for the width of the domain walls, this definition can describe both MPDs and conventional ferroelectric domains. Using this definition, it is clear that, although the domain walls move and the area of the domains change, the sign of the polarization inside a domain never changes. Thus, it is clear that the MPDs in 3R-stacked homo-bilayers are \textit{not} ferroelectric in a static homo-bilayer or via lattice relaxation, as claimed in Refs.~\onlinecite{ferreira2021weak,enaldiev2021piezoelectric,enaldiev2021scalable}. An inversion of the signs of the polarization in the domains may be possible in homo-bilayers via van der Waals sliding. Currently, this is the only known mechanism by which the MPDs could be classed as ferroelectric domains. However as mentioned previously, this has not yet been experimentally observed.

It has also been suggested that MPDs are antiferroelectric \cite{shen2019emergent,li2021extremely}. Although in 3R-stacked homo-bilayers the domains have opposite orientation and the polarization averages to zero, the classical definition of antiferroelectricity refers to anti-aligned adjacent \textit{dipoles}, not adjacent \textit{domains}, analogous to an antiferromagnet. A more general definition of antiferroelectricity is a material which can be made ferroelectric with an applied field \cite{rabe2013antiferroelectricity}, resulting in a double hysteresis loop, which has not been experimentally observed for moir\'e superlattices. Using either definition, it is not appropriate to classify moir\'e heterostructures as antiferroelectric.

\subsection{Final remarks}

The recent experimental discoveries of ferroelectric behavior have prompted a change in thought about moir\'e heterostructures, and it is clear that they should be considered polar materials. Despite this, there have been misconceptions about what exactly corresponds to ferroelectric behavior. The domains have incorrectly been called ferroelectric, and it is not clear whether electrically tunable lattice relaxation can lead to ferroelectricity without the sample being imperfect, or if van der Waals sliding in a twisted bilayer is possible. This highlights the importance of establishing a general theoretical understanding of moir\'e heterostructures as polar materials. In particular, great care should be taken when comparing moir\'e heterostructures to conventional ferroelectric materials, because the origin of polarization and domain formation in each case, summarized in Table \ref{table:ferro}, is completely different.

The are still many unanswered questions about the polar properties of moir\'e heterostructures. Electrically tunable lattice relaxation is generally well understood for homo-bilayers \cite{bennett2022electrically}, and has been described in detail for general moir\'e heterostructures in this paper. However, to my knowledge, a theoretical study of polarization switching via van der Waals sliding has yet to be done. Additionally, the role of electromechanical coupling in both electrically tunable lattice relaxation and van der Waals sliding is not well-known. Finally, a complete picture of MPDs in moir\'e heterostructures at general twist angles is missing. Addressing these questions will lead to a better understanding of the polar properties of moir\'e heterostructures, and may lead to practical advancements in the field of twistronics.

\section*{Acknowledgements} 

I would like to thank B.~Remez, E.~Artacho, M.~Stengel, E.~Bousquet and P.~Ghosez for helpful discussions. I would like to acknowledge funding from the EPSRC Centre for Doctoral Training in Computational Methods for Materials Science under grant number EP/L015552/1, St.~John's College (University of Cambridge), and the University of Li{\'e}ge under special funds for research (IPD-STEMA fellowship programme).\\

\section*{Appendix: Description of lattice relaxation calculations}

A complete overview of how lattice relaxation calculations are practically performed is presented here, i.e.~how to numerically minimize Eq.~\eqref{eq:V_tot_main} in an efficient manner. There are several viable approaches to perform lattice relaxation calculations: the first and perhaps most obvious option is to minimize the free energy of the model using variational methods and solve the resulting differential equations. However, solving nonlinear partial differential equations (PDEs) with periodic boundary conditions can be difficult, and in configuration space, increasing the moir\'e period (decreasing the twist angle) makes the PDEs more nonlinear and volatile, requiring the use of very fine grids. The second option is to make use of the periodic boundary conditions in configuration space by performing Fourier expansions of the displacement field and the stacking energy, then minimize the total energy with respect to the Fourier components of the strain field. However, a closed form solution cannot be obtained because the Fourier components of the stacking energy depend on the Fourier components of the displacement field: $\Vstack(\bvec{s} + \bvec{u}(\bvec{s}))$. One option is to update the Fourier components of the displacement field and total energy self-consistently until the total energy is minimized \cite{nam2017lattice}. Another option is to directly minimize the free energy with respect to Fourier components of the displacement field using optimization techniques \cite{carr2018relaxation,bennett2022electrically}, which is the method used in this paper.

\subsubsection{One-dimensional example}

To illustrate the method, consider a one-dimensional (1D) field theory, described by a free energy functional
\beq{eq:V_tot_1D}
V_{\text{tot}}[u] = \int_0^1 \Vtot(x,u(x),\partial_xu(x)) \dd x
\eec
where $\Vtot$ is nonlinear and periodic on the interval $[0,1]$. For example, if
\beq{}
\Vtot(x,u(x),\partial_xu(x)) = \frac{1}{2}B\lb \partial_x u\rb^2 + \V_0 \cos{(2\pi (x+u(x))}
\eec
then Eq.~\eqref{eq:V_tot_1D} would describe a Frenkel-Kontorova model, which has been used to model domain walls in moir\'e superlattices \cite{popov2011commensurate,lebedeva2016dislocations,lebedeva2019commensurate,bennett2022electrically}. We could minimize Eq.~\eqref{eq:V_tot_1D}: $\lb \partial_x\partial_{\partial_u} - \partial_u\rb V_{\text{tot}} = 0$, and solve the resulting ordinary differential equation (ODE) for $u(x)$, but this can become difficult if $\V$ is highly nonlinear, or for a higher number of dimensions. Instead, we can take advantage of the periodicity of the system by expanding the displacement field in a Fourier series:
\beq{eq:u_1D}
u(x) = \sum_k u_k e^{2\pi i k x}
\eec
after which, Eq.~\eqref{eq:V_tot_1D}, is no longer a functional of $u(x)$, but a function of $\{u_k\}$: $V_{\text{tot}}[u] \to V_{\text{tot}}(\{u_k\})$. Instead of minimizing Eq.~\eqref{eq:V_tot_1D} using variational methods and solving the resulting ODEs, we can simply minimize it by differentiating with respect to Fourier components $\{u_k\}$:
\beq{}
\bvec{G} \equiv \grad_{u_k}V_{\text{tot}} = 0, \qquad \grad_{u_k} \equiv \begin{bmatrix} \partial_{u_1} \\ \partial_{u_2} \\ \vdots \end{bmatrix}
\eep
The $k\textsuperscript{th}$ component of $\bvec{G}$ is
\beq{}
\partial_{u_k}V_{\text{tot}} = \int \partial_{u_k} \Vtot \dd x = \int \frac{\delta \Vtot}{\delta u} \lb\partial_{u_k} u\rb \dd x
\eec
where
\beq{}
\frac{\delta \Vtot}{\delta u} = \partial_x \frac{\partial \Vtot}{\partial(\partial u)} - \frac{\partial\Vtot}{\partial u} \equiv D
\eeq
is the functional derivative, and the partial derivative $\partial_{u_k}$ selects the $k\textsuperscript{th}$ Fourier basis function from $u$:
\beq{}
\partial_{u_k} u = e^{2\pi i k x}
\eep
Expanding the functional derivative in a Fourier series,
\beq{}
D = \sum_k D_k e^{2\pi i k x}
\eec
we get
\beq{}
\partial_{u_k}V_{\text{tot}} = \int D \ e^{2\pi i k x} \dd x = D_k
\eep
Thus the derivative of $V_{\text{tot}}$ with respect to the Fourier components $\{u_k\}$ is
\beq{}
\bvec{G} = \begin{bmatrix} D_1 \\ D_2 \\ \vdots \end{bmatrix}
\eec
where in principle there is an infinite number of components, but in practice they should be appropriately truncated to some number $\Nmax$ (all results should be well converged with respect to the number of components). We can now use root finding techniques to solve $\bvec{G} = 0$ for $\{u_k\}$. For example, using Newton's method, the $n\textsuperscript{th}$ approximation to the Fourier coefficients $\bvec{u}^n = \left[ u_1^n, u_2^n, \ldots , u_{\Nmax}^n\right]$ is
\beq{}
\bvec{u}^{n} = \bvec{u}^{n-1} - H^{-1}(\bvec{u}^{n-1})\bvec{G}(\bvec{u}^{n-1}) 
\eeq
where $H$ is the Hessian of $V_{\text{tot}}$. Although we have an analytic expression for $\bvec{G}$ for a given $\bvec{u}^n$, an analytic expression for $H$ cannot be obtained when $\Vtot$ is nonlinear. Thus in order to use Newton's method, $H$ would need to be calculated using finite differences, making the calculations more expensive. A better approach is to use a quasi-Newton method such as the Broyden–Fletcher–Goldfarb–Shanno algorithm (BFGS), where an approximation to the the inverse Hessian is used to choose the direction of the next iteration to update $\bvec{u}^n$.

\subsubsection{Lattice relaxation in moir\'e heterostructures}

The optimization techniques introduced in the previous section can be generalized to 2D and used to perform lattice relaxation calculations in moir\'e superlattices, or more generally any nonlinear field theory with crystal symmetries. In configuration space, Eq.~\eqref{eq:V_tot_main} becomes
\beq{eq:V_tot_main_config}
\begin{split}
V_{\text{tot}} = \frac{1}{\As}\int_{\As}\bigg[\bigg. \Vel(\grad\bvec{u}) +& \Vstack(\bvec{s}+\bvec{u}(\bvec{s})) \\
 +& \Velec(\bvec{s}+\bvec{u}(\bvec{s}),\Ef)\bigg.\bigg] \dd\bvec{s}
\end{split}
\eep
For small twist angles, the elastic energy in configuration space is, explicitly \cite{carr2018relaxation,bennett2022electrically},
\beq{eq:V_elastic_cartesian}
\begin{split}
\Vel =& \frac{\th^2}{2} \bigg[\bigg. B\lb \partial_{s_x}u_{s_y} - \partial_{s_y}u_{s_x}\rb^2 + \\ 
&\mu \lb \lb \partial_{s_x}u_{s_y} + \partial_{s_y}u_{s_x} \rb^2 
+ \lb \partial_{s_x}u_{s_x} - \partial_{s_x}u_{s_x}\rb^2 \rb \bigg.\bigg]
\end{split}
\eeq
where $B$ is the bulk modulus and $\mu$ is the shear modulus, both multiplied by the in-plane area and thus in units of energy per unit cell.

\begin{figure}[h!]
\centering
\includegraphics[width=0.7\columnwidth]{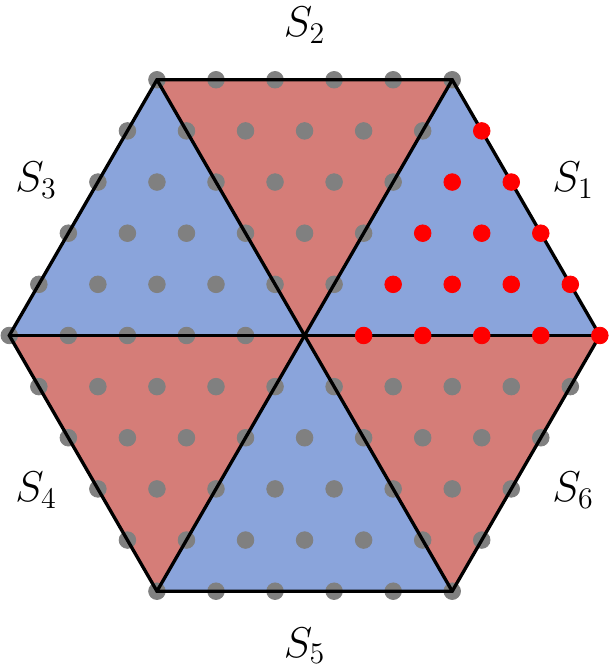}
\caption{BZ containing the $\bvec{k}$-vectors over which the displacement field and total energy density are expanded. The BZ is divded into six sectors, with $S_{2n+1}$ (light blue) and $S_{2n}$ (light red) being related by $\Ct$ rotations. The expansion over the entire BZ is reduced to an expansion over the vectors in $S_1$, highlighted in red, using the $\Ct$ rotation symmetry. }
\label{fig:BZ}
\end{figure}

We write the displacement field as a Fourier series expansion,
\beq{}
 \bvec{u}(\bvec{s}) = \sum_{\bvec{k}} \bvec{u}_{\bvec{k}} e^{i\bvec{k}\cdot\bvec{s}}
\eec
over the reciprocal lattice vectors $\bvec{k}$ in the Brillouin zone (BZ) of the configuration space unit cell, see Fig.~\ref{fig:BZ}. The number of independent $\bvec{k}$ vectors can be greatly reduced using crystal symmetries of the system, which for the bilayers considered in this paper is a $\Ct$ rotation symmetry. There is a mirror plane along the diagonal in configuration space which could be used to further reduce the number of independent $\bvec{k}$-vectors, but it is not used here. Scalar fields $\p$ and vector fields $\bvec{u}$ transform as
\beq{}
\begin{split}
\p_{\Ct\bvec{k}} & = \p_{\bvec{k}} \\
\bvec{u}_{\Ct\bvec{k}} & = \Ct\bvec{u}_{\bvec{k}}  \\
\end{split}
\eec
respectively. We split the sum over $\bvec{k}$ into 6 individual sectors $S_1,\ldots,S_6$, where the corresponding vectors in sectors $S_{2n}$ are equivalent via $\Ct$ rotations, as are the vectors in sectors $S_{2n+1}$, see Fig.~\ref{fig:BZ}. Thus, we can write
\beq{eq:u_full}
\begin{split}
 \bvec{u}(\bvec{s}) &= \sum_{n=1}^{6}\sum_{\bvec{k}}^{S_n} \bvec{u}_{\bvec{k}} e^{i\bvec{k}\cdot\bvec{s}} \\
 &= \sum_{n=1}^{3}\sum_{\bvec{k}}^{S_1} \Ct^{n-1}\bvec{u}_{\bvec{k}} e^{i\Ct^{n-1}\bvec{k}\cdot\bvec{s}} + 
 \sum_{n=1}^{3}\sum_{\bvec{k}}^{S_4} \Ct^{n-1}\bvec{u}_{\bvec{k}} e^{i\Ct^{n-1}\bvec{k}\cdot\bvec{s}}\\
 &= \sum_{n=1}^{3}\sum_{\bvec{k}}^{S_1} \lb \Ct^{n-1}\bvec{u}_{\bvec{k}} e^{i\Ct^{n-1}\bvec{k}\cdot\bvec{s}} + \Ct^{n-1}\bvec{u}_{-\bvec{k}} e^{-i\Ct^{n-1}\bvec{k}\cdot\bvec{s}} \rb\\
  &= \sum_{n=1}^{3}\sum_{\bvec{k}}^{S_1} \Ct^{n-1}\Big(\Big. \lb \bvec{u}_{\bvec{k}} + \bvec{u}_{-\bvec{k}}\rb \cos{\lb\Ct^{n-1}\bvec{k}\cdot\bvec{s}\rb} \\
  & \qquad\qquad\quad + i\lb \bvec{u}_{\bvec{k}} - \bvec{u}_{-\bvec{k}}\rb \sin{\lb\Ct^{n-1}\bvec{k}\cdot\bvec{s}\rb}  \Big.\Big)
 \end{split}
\eeq
Going from the first line to the second line, we use the $\Ct$ rotation symmetry to relate the Fourier components of equivalent $\bvec{k}$-vectors in sectors $S_{2n}$ and $S_{2n+1}$. The third line is arrived at by noting that $S_1$ and $S_4$ are related by $\bvec{k}\to -\bvec{k}$, although the Fourier components $u_{\bvec{k}}$ and $u_{-\bvec{k}}$ are independent unless there is an additional inversion symmetry. Finally, we rewrite the complex exponentials in terms of sine and cosine basis functions. In general, the displacement field has both even and odd terms, and it is convenient rewrite the Fourier components as
\beq{}
\begin{split}
\bvec{u}^{\text{even}}_{\bvec{k}} &= \bvec{u}_{\bvec{k}} + \bvec{u}_{-\bvec{k}} \\
\bvec{u}^{\text{odd}}_{\bvec{k}} &= i\lb \bvec{u}_{\bvec{k}} - \bvec{u}_{-\bvec{k}}\rb \\
\end{split}
\eeq
If there is also an inversion symmetry, giving the system an overall $\Cs$ rotation symmetry, then $\bvec{u}_{\bvec{k}} = - \bvec{u}_{-\bvec{k}}$, i.e.~the cosine terms vanish, and the displacement is purely odd:
\beq{eq:u_odd}
 \bvec{u}(\bvec{s}) = \sum_{n=1}^{3}\sum_{\bvec{k}}^{S_1} \Ct^{n-1}\bvec{u}_{\bvec{k}} \sin{\lb\Ct^{n-1}\bvec{k}\cdot\bvec{s}\rb}
\eec
where we have dropped the `odd' superscript. This is the case for 3R-stacked homo-bilayers at zero electric field, although not for 2H-stacked homo-bilayers or any hetero-bilayers. In the interest of simplicity, we consider a system with $\Cs$ rotation symmetry and therefore a displacement field given by Eq.~\eqref{eq:u_odd}, for the rest of this section. The generalization to a system with $\Ct$ rotation symmetry, and hence a displacement field given by Eq.~\eqref{eq:u_full} is obtained following the same methodology since the even and odd terms are linearly independent, although the algebra is more involved.

It is also convenient to work in reduced coordinates, i.e.~in terms of the lattice vectors $\bvec{a}_1 = \begin{bmatrix}1 \\ 0\end{bmatrix}$ and $\bvec{a}_2 = \begin{bmatrix}1/2 \\ \sqrt{3}/2\end{bmatrix}$, achieved by the transformation:
\beq{}
\begin{gathered}
\bvec{s} = g \bvec{x} \leftrightarrow \bvec{x} = g^{-1}\bvec{x}\\
g = \begin{bmatrix}1 & 1/2\\ 0 & \sqrt{3}/2\end{bmatrix}
\end{gathered}
\eec
In reduced coordinates, the unit vectors are non-orthogonal, but the unit cell is a square, making numerical integrations more efficient. The displacement field transforms as ${\bvec{u}(\bvec{s}) \to g \bvec{u}(g\bvec{x})}$, and the Fourier components transform as $\bvec{u}_{\bvec{k}} \to g \bvec{u}_{\bvec{k}}$. The exponentials and trigonometric functions transform as
\beq{}
\begin{split}
\exp{\lb i \bvec{k}\cdot\bvec{s}\rb} &= \exp{\lb i \bvec{k}\cdot\lb g\cdot g^{-1}\rb\bvec{s}\rb} \\
& = \exp{\lb i \bvec{k}\cdot g\bvec{x}\rb} \\
& = \exp{\lb i g^T\bvec{k}\cdot \bvec{x}\rb}
\end{split}
\eep
Thus, Eq.~\eqref{eq:u_odd} becomes
\beq{}
 \bvec{u}(\bvec{x}) = \sum_{n=1}^{3}\sum_{\bvec{k}}^{S_1} g^{-1}\Ct^{n-1}\bvec{u}_{\bvec{k}} \sin{\lb g^T\Ct^{n-1}\bvec{k}\cdot\bvec{x}\rb}
\eep
Note that the factor of $g^{-1}$ cannot be absorbed into the Fourier components because the $\Ct$ rotation symmetry is not preserved after the transformation to reduced coordinates. 

While $\Vstack$ and $\Velec$ are naturally obtained in reduced coordinates from first-principles calculations by using Eq.~\eqref{eq:fourier_fitting}, the elastic energy in Eq.~\eqref{eq:V_elastic} must be transformed from Cartesian to reduced coordinates:
\beq{eq:V_elastic}
\begin{split}
\Vel &= \frac{\th^2}{2} \bigg[\bigg. B\lb \partial_{x}u_{y} - \partial_{y}u_{x}\rb^2 \\
&+ \mu \lb \frac{4}{3}\lb \partial_{x}u_{y} + \partial_{y}u_{x} \rb^2 + \lb  \partial_{x}u_{x} - \partial_{y}u_{y}\rb^2 \rb \bigg.\bigg]
\end{split}
\eep\vskip 0.2in
Fortunately, we can write $\Vel$ exactly in terms of the Fourier components $\{\bvec{u}_{\bvec{k}}\}$. In order to do this, we need the derivatives of the displacement field:

\beq{eq:u_deriv}
\begin{split}
\partial_x \bvec{u} &= \sum_{n=1}^{3}\sum_{\bvec{k}}^{S_1}
\lb g^T\Ct^{n-1}\bvec{k}\rb_x\lb\Ct^{n-1}\bvec{u}_{\bvec{k}}\rb \cos{\lb g^T\Ct^{n-1}\bvec{k}\cdot\bvec{x}\rb}\\
\partial_y \bvec{u} & = \sum_{n=1}^{3}\sum_{\bvec{k}}^{S_1}
\lb g^T\Ct^{n-1}\bvec{k}\rb_y\lb\Ct^{n-1}\bvec{u}_{\bvec{k}}\rb \cos{\lb g^T\Ct^{n-1}\bvec{k}\cdot\bvec{x}\rb}\\
\end{split}
\eep
Inserting Eq.~\eqref{eq:u_deriv} into Eq.~\eqref{eq:V_elastic} we obtain, after some algebra,

\begin{widetext}
\beq{eq:V_elastic_exact}
\begin{split}
\mathcal{V}_{\text{elastic}} = \frac{1}{4}\th^2 \sum_{\bvec{k}}\sum_{n=1}^3 & B\lb \lb g^T\Ct^{n-1}\bvec{k}\rb_x  \lb g^{-1}\Ct^{n-1}\bvec{u}_{\bvec{k}}\rb_y - \lb g^T\Ct^{n-1}\bvec{k}\rb_y  \lb g^{-1}\Ct^{n-1}\bvec{u}_{\bvec{k}}\rb_x \rb^2 \\
&+ \mu \lb \frac{4}{3}\lb \lb g^T\Ct^{n-1}\bvec{k}\rb_x  \lb g^{-1}\Ct^{n-1}\bvec{u}_{\bvec{k}}\rb_y + \lb g^T\Ct^{n-1}\bvec{k}\rb_y  \lb g^{-1}\Ct^{n-1}\bvec{u}_{\bvec{k}}\rb_x \rb^2 \right. \\
&\left.+ \lb \lb g^T\Ct^{n-1}\bvec{k}\rb_x  \lb g^{-1}\Ct^{n-1}\bvec{u}_{\bvec{k}}\rb_x - \lb g^T\Ct^{n-1}\bvec{k}\rb_y  \lb g^{-1}\Ct^{n-1}\bvec{u}_{\bvec{k}}\rb_y \rb^2 \rb
\end{split}
\eec
\end{widetext}
where we use the orthogonality relations,
\beq{}
\begin{split}
\int \sin{(g^T \bvec{k}\cdot\bvec{x})}\sin{(g^T\bvec{k}'\cdot\bvec{x})}\dd{\bvec{x}} & = \frac{1}{2}\d_{\bvec{k},\bvec{k}'} \\
\int \cos{(g^T \bvec{k}\cdot\bvec{x})}\cos{(g^T\bvec{k}'\cdot\bvec{x})}\dd{\bvec{x}} & = \frac{1}{2}\d_{\bvec{k},\bvec{k}'}
\end{split}
\eec
to perform the integration. When the $\Cs$ rotation symmetry is reduced to $\Ct$, and the displacement field is given by Eq.~\eqref{eq:u_full}, Eqs.~\eqref{eq:u_deriv} and \eqref{eq:V_elastic_exact} will contain equivalent contributions from $\{\bvec{u}_{\bvec{k}}^{\text{even}}\}$.

Eq.~\eqref{eq:V_elastic_exact} is an exact expression of Eq.~\eqref{eq:V_elastic_cartesian} in terms of Fourier harmonics, meaning the elastic energy doesn't need to be calculated numerically, which results in a considerable reduction in computational effort. Unfortunately, the stacking energy cannot be obtained analytically since, as mentioned previously, the Fourier components of $\Vstack$ are functions of $\{\bvec{u}_{\bvec{k}}\}$.

Now we must obtain an analytic expression for the gradient as we did in the 1D example. First, we truncate the expansion so that a suitable number of shells are included, containing $\Nmax$ vectors in total. For example, in Fig.~\ref{fig:BZ}, the first 5 shells are shown, containing $\Nmax = 15$ vectors in $S_1$. The gradient is then a $2\times \Nmax$ array,
\beq{eq:G_initial}
\begin{split}
G &\equiv \nabla_{\bvec{u}_{\bvec{k}}} V_{\text{tot}} = \nabla_{\bvec{u}_{\bvec{k}}}\int \Vtot \ \dd\bvec{x} = \int \frac{\delta\Vtot}{\delta\bvec{u}} \nabla_{\bvec{u}_{\bvec{k}}} \bvec{u}\ \dd\bvec{x} \\[10pt]
&= \int \begin{bmatrix} D_x\partial_{u_{x,1}} u_x & D_y\partial_{u_{y,1}} u_y\\ \vdots & \vdots \\ D_x\partial_{u_{x,Nmax}} u_x & D_y\partial_{u_{y,Nmax}} u_y \end{bmatrix} \dd{\bvec{x}}
\end{split}
\eec
where the components of $\bvec{D}$ are the functional derivatives of $\Vtot$ with respect to the components of $\bvec{u}$:
\beq{}
\begin{split}
\bvec{D} &= \begin{bmatrix} D_x \\ D_y\end{bmatrix}\\
D_x & \equiv \frac{\delta \Vtot}{\delta u_x}\\
D_y & \equiv \frac{\delta \Vtot}{\delta u_y}
\end{split}
\eep
Eq.~\eqref{eq:G_initial} is general, and we would like to write the gradient in terms of the reduced set of $\bvec{k}$-vectors in $S_1$. First, note that 
\beq{}
\begin{split}
\Ct\partial_{u_{k_x}} \bvec{u}_{\bvec{k}} &= \Ct\begin{bmatrix} 1 \\ 0\end{bmatrix} = \begin{bmatrix} -1/2 \\ \sqrt{3}/2\end{bmatrix} \\
\partial_{u_{k_x}} \Ct\bvec{u}_{\bvec{k}} &= \partial_{u_{k_x}}\frac{1}{2}\begin{bmatrix} -u_{k_x} -\sqrt{3}u_{k_x} \\ \sqrt{3}u_{k_x} - u_{k_y}\end{bmatrix} = \begin{bmatrix} -1/2 \\ \sqrt{3}/2\end{bmatrix}
\end{split}
\eeq
i.e.~the derivative commutes with the $\Ct$ rotation operator (as well as the metric tensor). Now, the derivatives of $\bvec{u}$ with respect to the basis functions are
\beq{eq:f_k_x}
\begin{split}
\partial_{u_{k_x}}\bvec{u} & = \sum_{n=1}^3\sum_{\bvec{k}}^{S_1} \partial_{u_{k_x}}g^{-1}\Ct^{n-1}\bvec{u}_{\bvec{k}} \sin{\lb g^T\Ct^{n-1}\bvec{k}\cdot\bvec{x}\rb}\\
 & = \sum_{n=1}^3\sum_{\bvec{k}}^{S_1} g^{-1}\Ct^{n-1}\partial_{u_{k_x}}\bvec{u}_{\bvec{k}} \sin{\lb  g^T\Ct^{n-1}\bvec{k}\cdot\bvec{x}\rb}\\
  & = \sum_{n=1}^3\sum_{\bvec{k}}^{S_1} \sin{\lb g^T\Ct^{n-1}\bvec{k}\cdot\bvec{x}\rb} g^{-1}\Ct^{n-1}\hat{\bvec{x}} \equiv \bvec{f}_{\bvec{k}}^x(\bvec{x})
\end{split}
\eec
and similarly for $y$:
\beq{eq:f_k_y}
\partial_{u_{k_y}}\bvec{u} = \sum_{n=1}^3\sum_{\bvec{k}}^{S_1} \sin{\lb g^T\Ct^{n-1}\bvec{k}\cdot\bvec{x}\rb} g^{-1}\Ct^{n-1}\hat{\bvec{y}} \equiv \bvec{f}_{\bvec{k}}^y(\bvec{x})
\eec
where $\hat{\bvec{x}} = \begin{bmatrix} 1 \\ 0 \end{bmatrix}$ and $\hat{\bvec{y}} = \begin{bmatrix} 0 \\ 1 \end{bmatrix}$. Re-writing the gradient in terms of Eqs.~\eqref{eq:f_k_x} and \eqref{eq:f_k_y}, we get
\beq{}
\begin{split}
\partial_{u_{k_x}}V_{\text{tot}} &= \partial_{u_{k_x}}\int \Vtot(x+u_x, y+u_y)\dd\bvec{x}\\
& = \int \lb\frac{\delta\Vtot}{\delta u_x}\partial_{u_{k_x}}u_x + \frac{\delta \Vtot}{\delta u_y}\partial_{u_{k_x}}u_y\rb\dd\bvec{x} \\
& = \int \bvec{D}\cdot \bvec{f}_{\bvec{k}}^x(\bvec{x})\dd \bvec{x}
\end{split}
\eeq
for the $x$ component, and
\beq{}
\partial_{u_{k_y}}V_{\text{tot}}  = \int \bvec{D}\cdot \bvec{f}_{\bvec{k}}^y(\bvec{x})\dd \bvec{x}
\eeq
for the $y$ component. Thus, the gradient becomes
\beq{}
G = \int \begin{bmatrix}  \bvec{D}\cdot \bvec{f}^x_{\bvec{k}_1} & \bvec{D}\cdot \bvec{f}^y_{\bvec{k}_1}\\ \vdots & \vdots \\ \bvec{D}\cdot \bvec{f}^x_{\bvec{k}_{\Nmax}} & \bvec{D}\cdot \bvec{f}^y_{\bvec{k}_{\Nmax}}\end{bmatrix}\dd\bvec{x}
\eeq
The functional derivatives are, explicitly, 

\beq{}\resizebox{\columnwidth}{!}{$
\bvec{D} =
\begin{bmatrix}
 \partial_x\mathcal{V}_{\text{stack}} - \frac{1}{3}\theta^2 \lb3\mu\partial_x^2u_x -(3B-\mu)\partial_x\partial_yu_y + (3B+4\mu)\partial_y^2u_x\rb\\[10pt]
 \partial_y\mathcal{V}_{\text{stack}} - \frac{1}{3}\theta^2 \lb3\mu\partial_y^2u_y -(3B-\mu)\partial_x\partial_yu_x + (3B+4\mu)\partial_x^2u_y\rb\\
\end{bmatrix}
$}
\eeq

which can be verified using e.g.~the ${\text{\sc VariationalMethods}}$ package in {\sc Mathematica}.

The first term in $\bvec{D}\cdot \bvec{f}^j_{\bvec{k}}$ is the projection of the derivtatives of $\mathcal{\Vstack}$ onto the basis function $\bvec{f}^j_{\bvec{k}}$. These must be obtained with numerical integration, which may be expensive. Fortunately, rest of the terms, i.e.~the elastic energy, is exact and can be calculated for free for any $\bvec{u}_{\bvec{k}}$. Writing out the elastic part of the functional derivative, and defining
\begin{widetext}
\beq{}
\begin{split}
\bvec{\bar{D}}_{\text{el},\bvec{k}} &\equiv \sum_{n=1}^3\sum_{\bvec{k}}^{S_1} \int \bvec{D} \sin{\lb g^T\Ct^{n-1}\bvec{k}\cdot\bvec{x}\rb} \dd\bvec{x} \\
& =-\frac{1}{3}\theta^2\sum_{n=1}^3 \begin{bmatrix} 3\mu \lb \Ct^{n-1}\bvec{k}\rb_x^2\lb \Ct^{n-1}\bvec{u}\rb_x - (3B-\mu) \lb \Ct^{n-1}\bvec{k}\rb_x\lb \Ct^{n-1}\bvec{k}\rb_y\lb \Ct^{n-1}\bvec{u}\rb_y + (3B+4\mu) \lb \Ct^{n-1}\bvec{k}\rb_y^2\lb \Ct^{n-1}\bvec{u}\rb_x\\[10pt]
3\mu \lb \Ct^{n-1}\bvec{k}\rb_y^2\lb \Ct^{n-1}\bvec{u}\rb_y - (3B-\mu) \lb \Ct^{n-1}\bvec{k}\rb_x\lb \Ct^{n-1}\bvec{k}\rb_y\lb \Ct^{n-1}\bvec{u}\rb_x + (3B+4\mu) \lb \Ct^{n-1}\bvec{k}\rb_x^2\lb \Ct^{n-1}\bvec{u}\rb_y \end{bmatrix} 
\end{split}
\eeq
\end{widetext}
the elastic part of the gradient is, after some algebra,
\beq{eq:G_elastic}\resizebox{\columnwidth}{!}{$
G_{\text{el}} = \frac{1}{2}\sum_{n=1}^3 \begin{bmatrix} \bvec{\bar{D}}_{\text{el},\bvec{k}_1} \cdot \lb g^{-1} \Ct^{n-1}\cdot\hat{\bvec{x}}\rb & \bvec{\bar{D}}_{\text{el},\bvec{k}_1} \cdot \lb g^{-1}\Ct^{n-1}\cdot\hat{\bvec{y}}\rb  \\
\vdots & \vdots \\
\bvec{\bar{D}}_{\text{el},\bvec{k}_{\Nmax}} \cdot \lb g^{-1}\Ct^{n-1}\cdot\hat{\bvec{x}}\rb & \bvec{\bar{D}}_{\text{el},\bvec{k}_{\Nmax}} \cdot \lb g^{-1}\Ct^{n-1}\cdot\hat{\bvec{y}}\rb
\end{bmatrix}
$}
\eeq
which can be verified by calculating the gradient of Eq.~\eqref{eq:V_elastic} numerically using finite differences.

\clearpage


%

\clearpage

\clearpage

\includepdf[pages={1}]{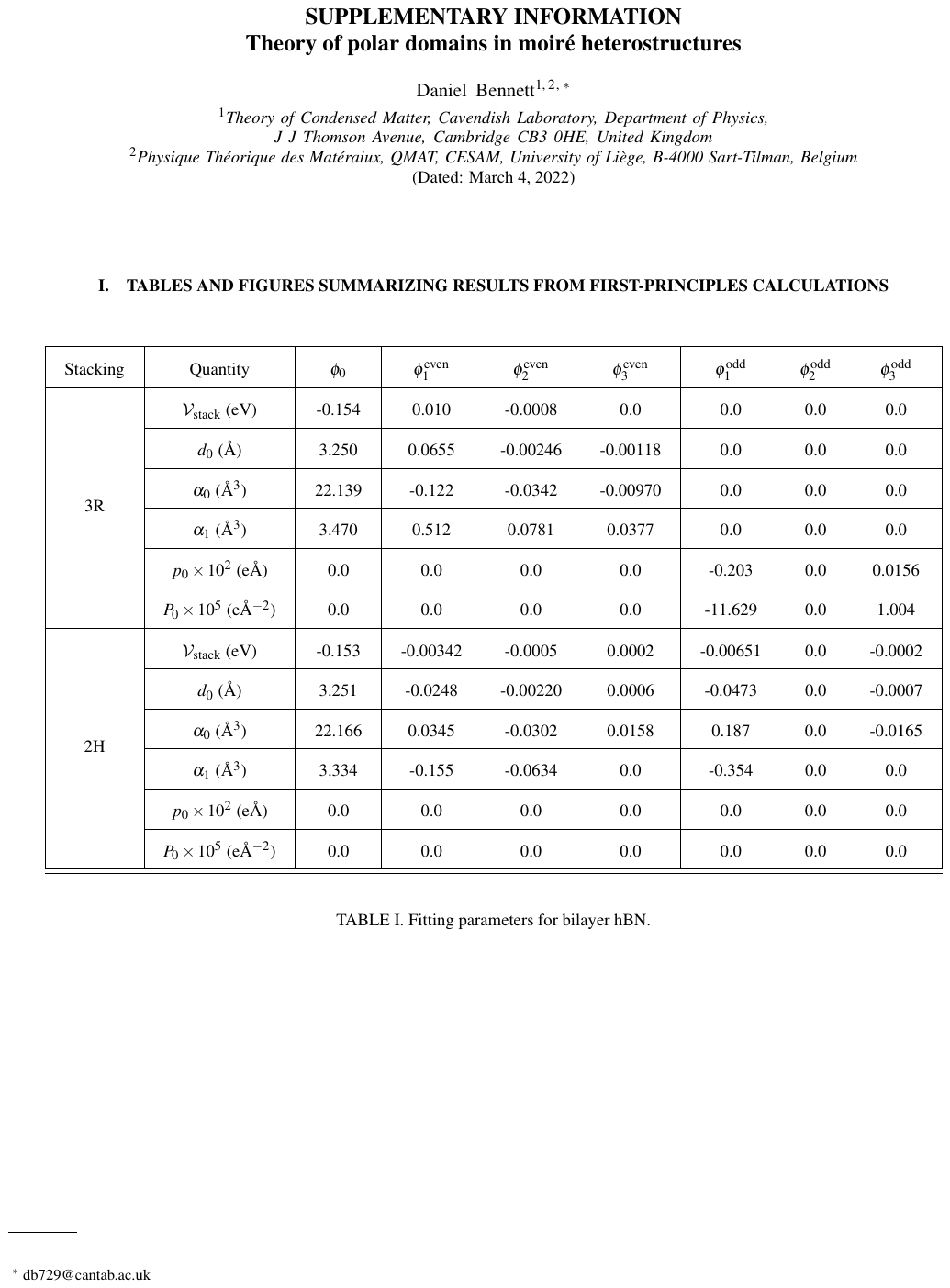}
\clearpage
\includepdf[pages={2}]{./SI.pdf}
\clearpage
\includepdf[pages={3}]{./SI.pdf}
\clearpage
\includepdf[pages={4}]{./SI.pdf}
\clearpage
\includepdf[pages={5}]{./SI.pdf}
\clearpage
\includepdf[pages={6}]{./SI.pdf}
\clearpage
\includepdf[pages={7}]{./SI.pdf}
\clearpage
\includepdf[pages={8}]{./SI.pdf}
\clearpage
\includepdf[pages={9}]{./SI.pdf}
\clearpage
\includepdf[pages={10}]{./SI.pdf}
\clearpage
\includepdf[pages={11}]{./SI.pdf}
\clearpage
\includepdf[pages={12}]{./SI.pdf}
\clearpage
\includepdf[pages={13}]{./SI.pdf}
\clearpage
\includepdf[pages={14}]{./SI.pdf}
\clearpage
\includepdf[pages={15}]{./SI.pdf}
\clearpage
\includepdf[pages={16}]{./SI.pdf}
\clearpage
\includepdf[pages={17}]{./SI.pdf}
\clearpage
\includepdf[pages={18}]{./SI.pdf}
\clearpage
\includepdf[pages={19}]{./SI.pdf}
\clearpage
\includepdf[pages={20}]{./SI.pdf}
\clearpage
\includepdf[pages={21}]{./SI.pdf}
\clearpage
\includepdf[pages={22}]{./SI.pdf}
\clearpage
\includepdf[pages={23}]{./SI.pdf}

\end{document}